\documentclass[aps,prl,twocolumn,10pt,showpacs,superscriptaddress]{revtex4-1}
\usepackage{amsmath,amssymb,graphicx,epsfig}
\usepackage{color}
\usepackage{mathtools}
\usepackage{mathrsfs}
 
\DeclarePairedDelimiter\ket{\lvert}{\rangle} 
\DeclarePairedDelimiterX\braket[2]{\langle}{\rangle}{#1 \delimsize\vert #2}

\begin{document}
\title{Stochastic Approach to Non-Equilibrium Quantum Spin Systems}
 \author{S. De Nicola}
\affiliation{Department of Physics, King's College London, Strand,
  London WC2R 2LS, United Kingdom} \author{B. Doyon}
\affiliation{Department of Mathematics, King's College London, Strand,
  London WC2R 2LS, United Kingdom} \author{M. J. Bhaseen}
\affiliation{Department of Physics, King's College London, Strand,
  London WC2R 2LS, United Kingdom}

\begin{abstract}
We investigate a stochastic approach to non-equilibrium quantum spin
systems based on recent insights linking quantum and classical
dynamics. Exploiting a sequence of exact transformations, quantum
expectation values can be recast as averages over classical stochastic
processes. We illustrate this approach for the quantum Ising model by
extracting the Loschmidt amplitude and the magnetization dynamics from
the numerical solution of stochastic differential equations. We show
that dynamical quantum phase transitions are accompanied by clear
signatures in the associated classical distribution functions,
including the presence of enhanced fluctuations.
We demonstrate that the
method is capable of handling integrable and non-integrable problems
in a unified framework, including those in higher dimensions.
\end{abstract}

\maketitle

Recent experimental advances in cold atomic gases
\cite{newtonsCradle,langen2015,lamporesi,weiler,weitenbergEtAl,bakrEtAl,reviewColdAtoms}
have catalyzed widespread interest in the non-equilibrium dynamics of
isolated quantum many-body systems \cite{eisertReview2015}. Questions
ranging from the nature of thermalization
\cite{deutsch91,srednicki94,rigol08,cramer08} to the growth of
entanglement following a quantum quench \cite{calabreseCardy2007} have
attracted considerable theoretical attention. In one dimension, the
availability of analytical techniques based on integrability has led
to fundamental insights into the role of conservation laws and the
Generalized Gibbs Ensemble (GGE) \cite{rigolPRL2007,polkovnikov2011,gogolinEisert2016,esslerFagotti2016}. 

In spite of these advances, much less is known about the behavior of non-integrable systems, where very few analytical techniques are
available. The situation is particularly challenging in higher dimensions, where
the rapid growth of the Hilbert space also stymies numerical
simulations, even in equilibrium. Recent progress 
includes the development of hydrodynamic approaches to non-equilibrium
steady states, based on macroscopic conservation laws and
thermodynamic equations of state \cite{BhaseenDoyonLucasSchalm2015,
  DoyonLucasSchalmBhaseen2015,bernardDoyon2016,castro-alvaredoDoyonYoshimura2017}.  Significant advances have also been
made using machine-learning algorithms
\cite{Carleo602,heylSchmitt2018} by exploiting novel representations of the
quantum wavefunction.

In this manuscript, we explore a rather different approach to
non-equilibrium systems based on an exact mapping between quantum spin
dynamics and classical stochastic processes
\cite{hoganChalker,galitski,ringelGritsev}.
By a sequence of exact transformations, stochastic differential
equations (SDEs) can be derived whose solutions
yield quantum expectation values. We show that this approach can be
turned into a viable tool for exploring quantum many-body dynamics in
both integrable and non-integrable settings, including higher
dimensions.

{\em Stochastic Formalism.}--- The method is readily illustrated by considering the quantum Hamiltonian
\begin{equation}\label{hamiltonian}
  \hat H=\sum_{ij}J_{ij}^{ab}\hat S_i^a\hat S_j^b+\sum_i h_i^a\hat S_i^a,
  \end{equation}
where the spin operators $\hat S^a_j$ on site $j$ obey the commutation
relations $[\hat S_j^a,\hat
  S_{j^\prime}^b]=i\delta_{jj^{\prime}}\epsilon^{abc}\hat S_j^c$ and
we set $\hbar=1$.  Here $J_{ij}^{ab}$ is the exchange interaction and
$h_i^a$ is an applied magnetic field with arbitrary orientation. The
dynamics of the model is governed by the time evolution operator
\begin{equation}
\hat U(t_f,t_i)={\mathbb T}\exp\left(-i\int_{t_i}^{t_f} \mathrm{d}t\,\hat
H(t)\right),
\end{equation}
between initial and final times $t_i$ and $t_f$, where $\hat H(t)$ can
be time-dependent and ${\mathbb T}$ denotes time ordering. The
operator $\hat U$ is non-trivial, due to the interactions in $\hat H$,
the non-commutativity of the spin operators, and the
time-ordering. However, $\hat U$ can be expressed in an alternative
form by means of a sequence of exact transformations
\cite{hoganChalker,galitski, ringelGritsev}. To begin with, the
interactions can be decoupled using Hubbard--Stratonovich
transformations \cite{hubbard,stratonovich} over auxiliary variables
$\phi_i^a$:
\begin{equation}
  \hat U = {\mathbb T}\int {\mathcal  D}\phi\,\exp \left(- i S - i \int_{t_i}^{t_f}  \mathrm{d}t \sum_j (\phi_j^a \hat  S_j^a + h_j^a \hat S_j^a) \right),
  \label{decoupled}
\end{equation}                  
where ${\mathcal D}\phi \equiv \prod_j \mathcal{D} \phi_j^a$ and the
normalization factors have been absorbed into the measure.
Eq.~(\ref{decoupled}) describes decoupled spins interacting with
stochastic magnetic fields $\phi_i^a$ governed by the Gaussian action
\begin{equation}
S = \frac{1}{4} \int_{t_i}^{t_f} \mathrm{d}t   \sum_{ij} (J^{-1})_{ij}^{ab}\phi^a_i \phi^b_j. 
\label{noiseAction}
\end{equation}  
  Equivalently,
\begin{equation}\label{stochasticFields}
  \hat U(t_f,t_i)=\big\langle {\mathbb T}\,
  e^{-i\int_{t_i}^{t_f} dt \sum_j\Phi_j^a(t)\hat S_j^a(t)}\big\rangle_\phi,
\end{equation}
where $\Phi_j^a\equiv \phi_j^a+h_j^a$ and the average $\langle\dots
\rangle_\phi$ is taken with the action in
Eq.~(\ref{noiseAction}). This can be further simplified as the
time-ordered exponential in Eq.~(\ref{stochasticFields}) can be
directly expressed as a group element
\cite{hoganChalker, galitski, ringelGritsev} using the
Wei--Norman--Kolokolov decomposition for $\mathrm{SU(2)}$ \cite{weiNorman, kolokolov}:
\begin{equation}
\hat U(t_f,t_i)=\big\langle \prod_j e^{\xi_j^+(t_f)\hat
  S_j^+}e^{\xi_j^z(t_f)\hat S_j^z}e^{\xi_j^-(t_f)\hat
  S_j^-}\big\rangle_\phi,
\end{equation}
where $\hat S_j^\pm=\hat S_j^x\pm i \hat S_j^y$. The coefficients $\xi_j^a$ are referred to as {\em disentangling variables} \cite{ringelGritsev} and are related to the original $\Phi_j^a$ via
\begin{subequations}
  \begin{align}
  i \dot\xi_j^+ & = \Phi_j^+ +\Phi_j^z \xi_j^+ -\Phi_i^- {\xi_j^+}^2,\\
  i \dot \xi_j^z & = \Phi_j^z -2 \Phi_j^- \xi_j^+ \ ,\\
  i \dot \xi_j^- & = \Phi_j^- \exp\xi_j^z \ ,
  \end{align}
  \label{disentangle}%
\end{subequations}
where $\xi^a_i(t_i)=0$. These equations are non-linear SDEs for the complex variables $\xi_j^a$, where the Hubbard--Stratonovich variables $\phi_j^a$ represent Gaussian noise. Indeed, the SDEs can be put in the canonical form \cite{kloeden}:
\begin{equation}
  \frac{d\xi_i^a}{dt}=A_i^a(\{\xi_i\})+\sum_{jb}B_{ij}^{ab}(\{\xi_i\})\bar\phi_j^b,
  \label{SDEs}
\end{equation}
where $A_i^a$ and $B_{ij}^{ab}$ are the drift and diffusion
coefficients respectively with $\{\xi_i\}=(\xi_i^z,\xi_i^\pm)$, and
$\bar\phi_j^b$ are delta-correlated white noise variables obtained by
diagonalizing the action in Eq.~(\ref{noiseAction}) \footnote{In order
  to diagonalize the action (\ref{noiseAction}), the matrix
  $(J_{ij}^{ab})^{-1}$ should be symmetric and invertible. For system
  sizes that are multiples of $4$ we add a constant to $\hat H$ to
  eliminate zero eigenvalues. This modifies the evolution of the stochastic variables but does not influence physical observables.}. These exact transformations allow one
to recast quantum dynamics in terms of SDEs, where quantum expectation
values are replaced by averages over classical processes.  This method
has been applied to the thermodynamics of a single cluster of quantum
spins \cite{hoganChalker} and to the dynamics of a single spin coupled
to a photonic waveguide \cite{ringelGritsev}.
Here, we show that this novel approach can be applied to both
integrable and non-integrable lattice spin models, including those in
higher dimensions. In general, the numerical solution of non-linear
SDEs using the Euler scheme may have divergent trajectories where
the stochastic variables, such as $\xi_j^+(t)$, grow without bound
\cite{HutzenthalerKloeden}. Throughout this manuscript we present results obtained from the non-divergent trajectories. In evaluating quantum expectation values we retain more than 99\% of the realizations at the stopping time. In plotting the associated classical variables for large system sizes, we typically retain more than 90\% of the trajectories.

\begin{figure}
  \includegraphics[trim={0.1cm 0.0cm 0.75cm
      0.3cm},clip,width=3.2in]{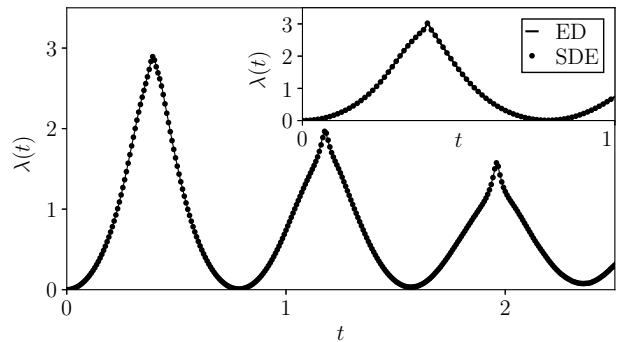}
  \caption{Loschmidt rate function $\lambda(t)$ for the $1$D quantum
    Ising model following a quantum quench from $\Gamma = 0$ to
    $\Gamma = 16\,\Gamma_c$, across the quantum critical point at
    $\Gamma_c=J/2$. The results obtained from the SDE approach (filled
    circles) are in excellent agreement with ED (solid line) for $N=7$
    spins. The results show clearly resolved peaks. The SDE results
    were obtained by averaging over $5 \times 10^5$ realizations of
    the stochastic process with a discretization time-step
    $dt=10^{-5}$. The inset shows the first Loschmidt peak for the
    same quench parameters and $N=14$. The SDE result was obtained as
    the average of $3.2 \times 10^6 $ trajectories with $dt=10^{-5}$.}
  \label{Fig:Loschmidt}
\end{figure}

\begin{figure}
  \includegraphics[trim={0.1cm 0.15cm 0.7cm
      0.3cm},clip,width=3.2in]{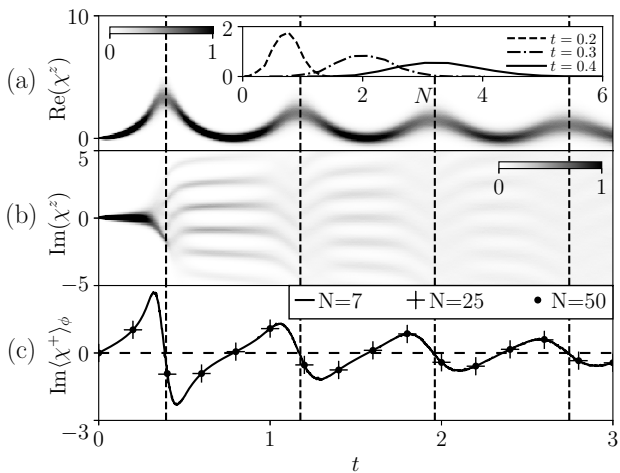}
  \caption{Time-evolution of the distribution of $\chi^z(t)\equiv
    N^{-1}\sum_j \xi_j^z(t)$ for the quantum Ising model following a
    quantum quench from $\Gamma=0$ to $\Gamma=16\Gamma_c$ with
    $N=7$. (a) The distribution of ${\rm Re}\,\chi^z(t)$ shows smooth
    maxima and increased fluctuations in the vicinity of the Loschmidt
    peaks (dashed lines at $t=0.39$, $1.18$, $1.96$, $2.75$ obtained
    by ED). Inset: the average value and width of the distribution of
    ${\rm Re}\,\chi^z(t)$ increases on approaching the Loschmidt peaks
    as illustrated for the first peak. (b) The distribution of ${\rm
      Im}\,\chi^z(t)$ also shows signatures in the vicinity of the
    Loschmidt peaks. (c) Time-evolution of ${\rm Im}\langle
    \chi^+(t)\rangle_\phi$ for $N=7$, $25$, and $50$. The zeros of
    ${\rm Im}\langle \chi^z(t)\rangle_\phi$ occur in proximity
    to the turning points of $\lambda(t)$. }
  \label{Fig:stochVariables}
\end{figure}

\begin{figure}
  \includegraphics[trim={0.1cm 0.2cm 0.3cm 0.3cm},clip,width= \linewidth]{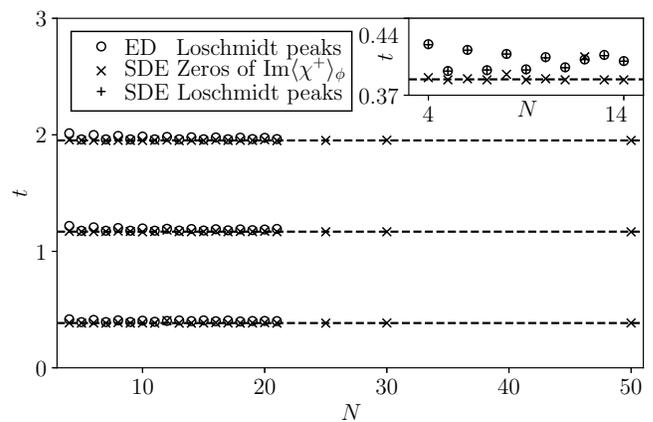}
  \caption{Characteristic times for the stationary points
    of ${\rm Re}\langle \chi^z(t)\rangle_\phi$ corresponding to the
    zeros of ${\rm Im}\langle\chi^+(t)\rangle_\phi$ following the
    quench considered in Fig.~\ref{Fig:stochVariables}. The times are
    in close proximity to the Loschmidt peaks and have little
    dependence on $N$. Inset: Comparison of the
    Loschmidt peak times obtained from the SDE approach by averaging
    the complete exponential in Eq.~(\ref{loschmidt}), including its
    real and imaginary parts and their correlations, and ED for
    different system sizes.}
  \label{Fig:times}
  \end{figure}

{\em Loschmidt Amplitude.}--- A natural quantity to study using the stochastic formalism is the Loschmidt amplitude $A(t)$, defined as the probability amplitude to return to an initial state $|\psi(0)\rangle$ after time $t$:
\begin{equation}
  A(t)=\langle \psi(0)|\hat U(t,0)|\psi(0)\rangle .
\end{equation}
In order to provide explicit results, we first examine the quantum Ising model in a transverse field $\Gamma$
\begin{equation}
  \hat H_{\rm I}= -J\sum_{j=1}^N \hat S_j^z\hat S_{j+1}^z - \Gamma \sum_{j=1}^N
  \hat S_j^x,
  \label{Ising}
\end{equation}
where $N$ is the number of lattice sites. We consider ferromagnetic
interactions $J>0$ and impose periodic boundary conditions; in the
numerical simulations below we set $J=1$ and measure time in units of
$J$. We take $|\psi(0)\rangle=\otimes_j\ket{\downarrow}_j\equiv
\ket{\Downarrow}$ with all spins down, corresponding to a
ferromagnetic initial state. For this initial state, the Loschmidt
amplitude is given by
\begin{equation}
  A(t)=\Big\langle \prod_{j=1}^N
  \exp\left(-\frac{\xi_j^z(t)}{2}\right)\Big\rangle_\phi,
  \label{loschmidt}
  \end{equation}
where the disentangling variables $\xi^z_j$ satisfy the SDEs
(\ref{disentangle}) with the appropriate model specific
coefficients. The amplitude (\ref{loschmidt}) can be obtained by
averaging over different realizations of the stochastic process. In
Fig.~\ref{Fig:Loschmidt} we plot the associated rate function
$\lambda(t)=-N^{-1}\ln |A(t)|^2$,
for unitary evolution in a non-zero transverse field. For quenches
across a quantum critical point, $\lambda(t)$ is known to exhibit
sharp peaks, corresponding to dynamical quantum phase transitions
(DQPTs) in the thermodynamic limit
\cite{HeylPolkovnikovKehrein,KarraschSchuricht,heyl2014}. Fig.~\ref{Fig:Loschmidt} shows that the SDE method is able to resolve these peaks for a quantum
quench across the critical point at $\Gamma_c=J/2$. The SDE results
are in excellent agreement with exact diagonalization (ED) results
obtained via the QuSpin package \cite{quSpin}.
Remarkably, the presence of the DQPTs is reflected in the
disentangling variables themselves. In Fig.~\ref{Fig:stochVariables}
we plot the time evolution of the distribution of $\chi^a(t)\equiv
N^{-1}\sum_{j}\xi_j^a(t)$ with $a=z$, as suggested by
Eq.~(\ref{loschmidt}).
It can be seen in Fig.~\ref{Fig:stochVariables}(a) that both the
average value and the width of the distribution of ${\rm
  Re}\,\chi^z(t)$ have smooth maxima in the vicinity of the DQPTs, as
further illustrated in the inset. Likewise, ${\rm Im}\,\chi^z(t)$
shows pronounced signatures close to the DQPTs, as indicated in
Fig.~\ref{Fig:stochVariables}(b); these features become less visible
with increasing $N$, and the overall phase of the argument of
Eq.~(\ref{loschmidt}) becomes uniformly distributed over $[-\pi,\pi]$
due to its scaling with $N$. Further insight into the location of the
DQPTs can be obtained from the SDEs. From Eq.~(\ref{disentangle}) it
can be seen that the turning points of ${\rm Re}\langle
\chi^z(t)\rangle_\phi$ are determined by the zeros of ${\rm Im}\langle
\chi^+(t)\rangle_\phi$ due to the exact relationship $\langle \dot
\chi^z(t)\rangle_\phi=-i\Gamma\langle \chi^+(t)\rangle_\phi$ for the
Ising SDEs. These zeros occur in close proximity to the DQPTs as shown
in Figs.~\ref{Fig:stochVariables}(c) and \ref{Fig:times}. The
expectation values $\langle\chi^z(t)\rangle_\phi$ and
$\langle\chi^+(t)\rangle_\phi$ and the characteristic times obtained
from the classical distribution functions show strikingly little
dependence on the system size, with results shown up to $N=50$. For
comparison, we show the exact locations of the Loschmidt maxima
obtained by averaging the complete exponential in
Eq.~(\ref{loschmidt}), including both the real and imaginary parts of
$\xi_j^z$ and their correlations; see inset of
Fig.~\ref{Fig:times}. The results are in very good agreement with ED.

\begin{figure}
\centering
\includegraphics[trim={0.3cm 0.5cm 1cm
      0.5cm},clip,width=3.2in]{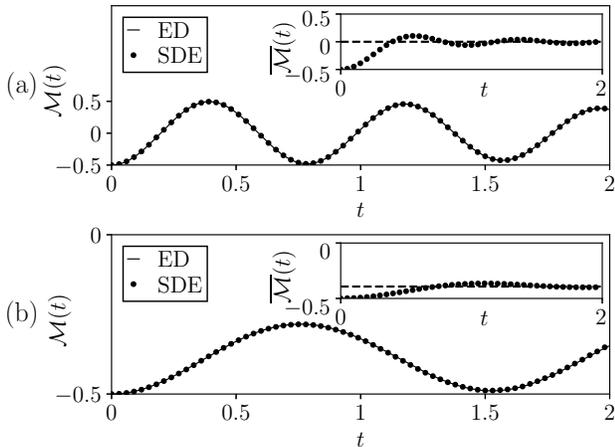}
\caption{(a) Time-evolution of $\mathcal{M}(t)$ for the quantum Ising
  model following a quantum quench from $\Gamma=0$ to $\Gamma =
  16\,\Gamma_c$. The results obtained from the SDE (full circles) are
  in excellent agreement with ED (solid line) for $N=3$. Inset:
  the time-averaged magnetization $\overline{\mathcal{M}}(t)$
  approaches zero at late times, as expected for the integrable
  case with $h=0$. The SDE results were obtained by averaging over $10^6$ realizations of the stochastic process with $dt=10^{-5}$.
  (b) $\mathcal{M}(t)$ for the non-integrable Ising model with $h=3J$,
  after a quench from $\Gamma=0$ to $\Gamma=2 J$. The results obtained
  from the SDEs are in agreement with ED for $N=3$. Inset: the
  time-averaged magnetization $\overline{\mathcal{M}}(t)$ approaches
  the thermal value calculated via ED (dashed line) at late times, as
  expected for the non-integrable case. The SDE results were obtained
  by averaging over $10^6$ realizations of the stochastic
  process with $dt=10^{-5}$.}
\label{Fig:MagInt}
\end{figure}

{\em Local Observables.}--- The stochastic approach can also be
applied to other physical observables including the 
magnetization. Following a quench from an initial state
$|\psi(0)\rangle$, the local magnetization evolves according to
\begin{equation}
  \langle \hat S_i^z(t)\rangle=\langle \psi(0)|\hat U^\dagger(t)\hat
  S_i^z\hat U(t)|\psi(0)\rangle.
\end{equation}
The forwards and backwards time-evolution operators can be decoupled by independent Hubbard--Stratonovich variables, $\phi_i^{a}$ and
$\tilde\phi_i^{a}$, with corresponding disentangling variables
$\xi_i^a(\phi)$ and $\tilde \xi_i^a(\tilde\phi)$. For a quantum quench starting in the ferromagnetic ground state $|\psi(0)\rangle=\ket{\Downarrow}$ with $\Gamma=0$, and time-evolving with $\Gamma\neq 0 $, one obtains
\begin{equation}
\langle \hat S_i^z(t)\rangle=\bigg\langle
f_i\left(\xi(t),\tilde\xi(t)\right)
\bigg\rangle_{\phi,\tilde\phi},
\label{magnetization}
\end{equation}
where $$f_i=-\frac{1}{2}e^{-\sum_j
  \tfrac{\xi_j^z+(\tilde\xi_j^z)^*}{2}}
[1-\xi_i^+(\tilde\xi_i^+)^*]\prod_{j\neq i}
[1+\xi_j^+(\tilde\xi_j^+)^*]. $$ In Fig.~\ref{Fig:MagInt}(a) we show
the time-evolution of the magnetization $\mathcal{M}(t)=
N^{-1}\sum\limits_{i=1}^N \langle \hat S_i^z(t)\rangle$
obtained from the stochastic average in Eq.~(\ref{magnetization}). The
results are in excellent agreement with ED for $N=3$. The use of
smaller system sizes enables us to reach longer time-scales in the
presence of two Hubbard--Stratonovich transformations. This allows us to verify that the time-integrated magnetization
$\overline{\mathcal{M}}(t)= 1/t \int_0^t
\mathrm{d}s\,\mathcal{M}(s)$ (inset) approaches zero at late times as
expected for the integrable Ising model in the absence of a
longitudinal magnetic field \cite{calabreseEsslerLetter2011}.
In Fig.~\ref{Fig:MagInt}(b) we show the dynamics of $\mathcal{M}(t)$
in the presence of a constant integrability-breaking longitudinal
field $h$, so that $\hat H=\hat H_{\rm I}+h\sum_j\hat S^z_j$. We
consider a quantum quench from the ferromagnetic ground state
$\ket{\Downarrow}$ with $\Gamma = 0$ and $h=0$ to $\Gamma = 2J$ and $h=3
J$. Again, the results are in excellent agreement with ED. In this
case, the time-averaged magnetization $\overline{\mathcal{M}}(t)$
approaches a non-vanishing expectation value as expected for
the non-integrable Ising model with $h\neq 0$ \cite{eisertReview2015}. The asymptotic result is consistent with the thermal expectation value obtained via ED.

\begin{figure}
  \includegraphics[trim={0.1cm 0.15cm 0.3cm
      0.1cm},clip,width=3.2in]{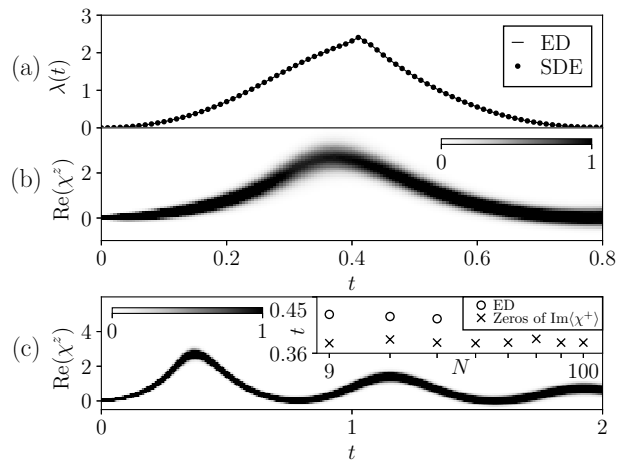}
  \caption{(a) Loschmidt rate function $\lambda(t)$ for the $2$D
    quantum Ising model following a quantum quench from $\Gamma=0$ to
    $\Gamma = 8 J$. The results obtained from the SDEs (filled
    circles) are in excellent agreement with ED (solid line) for a
    $3\times 5$ system. The results show sharp peaks in
    $\lambda(t)$ for quenches across the critical point at
    $\Gamma_c^{\rm 2D}\sim 1.523 J$. The SDE results were obtained by
    averaging over $2.5 \times 10^7$ stochastic realizations with $dt=10^{-5}$. (b) The corresponding distribution of
    ${\rm Re}\,\chi^z(t)$ for a $3\times 5$ system shows smooth maxima
    and increased fluctuations in the vicinity of the Loschmidt
    peaks. (c) Time-evolution of ${\rm Re}\,\chi^z(t)$ for a $10\times
    10$ spin system showing additional turning points. Inset: comparison of exact times of the first Loschmidt peak (circles) and zeros of Im$\langle \chi^+\rangle$ (crosses) for square lattices of size up to $N=100$ spins.}
  \label{Fig:Higher} 
\end{figure}

{\em Higher Dimensions.}--- A remarkable feature of the stochastic
approach is that it is not restricted to one-dimensional systems. To
illustrate this we examine the quantum Ising model in
$2+1$ dimensions:
\begin{equation}
  \hat H_{\rm I}^{\mathrm{2D}}= -J\sum_{\langle \mathbf{i} \mathbf{j} \rangle} \hat S_{\mathbf{i}}^z\hat S_{\mathbf{j}}^z - \Gamma \sum_{\mathbf{i}}
  \hat S_{\mathbf{i}}^x.
  \label{Ising2D}
\end{equation}
In Fig.~\ref{Fig:Higher}(a) we show $\lambda(t)$ following a
quench from $\Gamma=0$ to $\Gamma = 8 J$, across the 2D quantum
critical point at $\Gamma_c^{\mathrm{2D}} \sim 1.523 J$
\cite{pfeutyElliott1971,deJongh1998}. We initialize the system in the
ferromagnetic ground state $\ket{\Downarrow}$ and time evolve the 2D generalization of Eq.~(\ref{loschmidt}) using the SDEs in
Eq.~(\ref{SDEs}). The results are in excellent agreement with ED for a
$3 \times 5$ system.
The 2D results in Fig.~\ref{Fig:Higher}(a) show clear peaks in
$\lambda(t)$, as found for coupled continuum chains and for
classically tractable quenches from $\Gamma=\infty$ to $\Gamma=0$
\cite{jamesKonik2015,heyl2015}. Here, however, the SDE results apply
directly to the 2D quantum lattice model (\ref{Ising2D}), without continuum
approximations or assumptions of classical evolution. Moreover, the 2D
DQPTs are signalled once again by the presence of enhanced fluctuations
in the distribution of the disentangling variables, and the behavior
of their classical averages; see Fig.~\ref{Fig:Higher}(b). The
dynamics of these variables can be tracked to larger system sizes as
shown in Fig.~\ref{Fig:Higher}(c) for a $10\times 10$ system. This provides
a novel handle on the dynamics of higher-dimensional quantum many-body
systems. As found in 1D, the time evolution of the classical average
$\langle \chi^z\rangle_\phi$ and its turning points are strikingly
independent of $N$. This, together with the form of Eq.~(\ref{loschmidt}),
suggests the possibility of developing a classical large deviation
approach to quantum dynamics in future work.

{\em Conclusions.}--- In this manuscript we have explored the dynamics
of non-equilibrium quantum spin systems via an exact mapping to
classical stochastic processes. We have shown that this approach can
handle the dynamics of integrable and non-integrable systems,
including those in higher dimensions.
This novel approach provides a valuable handle on challenging problems
out of equilibrium and provides fundamental links between quantum and
classical dynamics. There are many directions for future research
including comparison with tensor network and machine learning
approaches, and the development of enhanced numerical sampling
techniques for the SDEs.

{\em Acknowledgements.}--- We acknowledge helpful conversations with
Samuel Begg, George Booth, Andrew Green, Vladimir Gritsev and Lev
Kantorovich. MJB is indebted to John Chalker for early discussions on
the Hubbard--Stratonovich and stochastic approaches to quantum
dynamics. SD acknowledges funding from the EPSRC Centre for Doctoral
Training in Cross-Disciplinary Approaches to Non-Equilibrium Systems
(CANES) under grant EP/L015854/1.  MJB, BD and SD thank the Centre for
Non-Equilibrium Science (CNES) and the Thomas Young Centre (TYC). We
are grateful to the UK Materials and Molecular Modelling Hub for
computational resources, which is partially funded by EPSRC
(EP/P020194/1). We acknowledge computer time on the Rosalind High
Performance Computer Cluster.

\end{document}